\documentclass[aps,pre,showpacs,showkeys,groupedaddress,preprint]{revtex4-1}  
\usepackage[cp1251]{inputenc}
\usepackage{graphicx}
\usepackage[english]{babel}
\usepackage{amsmath}
\begin{document}

\title{Generalized diffusion equation with fractional derivatives within Renyi statistics}

\author{P. Kostrobij$^{1}$, B. Markovych$^{1}$, O. Viznovych$^{1}$, M. Tokarchuk$^{1,2}$}

\affiliation{$^1$Lviv Polytechnic National University, 12 Bandera Str., 79013 Lviv, Ukraine\\
$^2$Institute for Condensed Matter Physics of NAS of Ukraine, 1 Svientsitskii Str.,
79011 Lviv, Ukraine}


\begin{abstract}
 By using the Zubarev nonequilibrium statistical operator method,
 and the Liouville equation with fractional derivatives,
 a generalized diffusion equation with fractional derivatives is obtained within the Renyi statistics.
 Averaging in generalized diffusion coefficient is performed with a power distribution with the Renyi parameter~$q$.
\end{abstract}

\keywords{generalized diffusion equation, nonequilibrium statistical operator, Renyi statistics, anomalous diffusion}

\maketitle

 \section{Introduction}

 The fractional derivatives and integrals~\cite{Oldham2006} are widely used
 to study anomalous diffusion in porous media~\cite{Uchaikin2008500,Sahimi1998213,Korosak20071,Hobbie2007,Metzler20001,Hilfer19951475,Hilfer20003914,Hilfer2000,Kosztolowicz2005041105,Kosztolowicz2015P10021,Bisquert20002287,Bisquert2001112,Kosztolowicz2009055004,Pyanylo201484},
 in disordered systems~\cite{Berkowitz19985858,Bouchaud1990127},
 in plasma physics~\cite{Balescu19954807,Tribeche2011103702,Gong2012023704,Carreras20015096,Tarasov2005082106,Tarasov2006052107},
 in turbulent~\cite{Monin1955256,Klimontovich2002,Zaslavsky2002461},
 kinetic and reaction-diffusion processes~\cite{Zaslavsky2002461,Zaslavsky1994110,Saichev1997753,Zaslavsky2004128,Nigmatullin2006282,Chechkin200278,Gafiychuk2007055201,Kosztolowicz2008066103,Shkilev20131066},
 in quantum mechanics~\cite{Laskin2000780,Laskin20003135,Laskin2000298,Laskin2002056108,Naber20043339}, etc.~\cite{Uchaikin2008500,Uchaikin20131074}

 Experimental data on the different processes of anomalous diffusion show that
 not only the law distribution,
 but also form of diffusion package is significantly different from the normal diffusion~\cite{Uchaikin2008500,Bouchaud1990127,Zaslavsky2002461,Uchaikin20131074}.
 Approaches with variable diffusion coefficients \cite{O'Shaughnessy1985455}, on the basis of a degree correlation of fractional order \cite{Mandelbrot1982}, fractional derivatives \cite{Monin1955256,Klimontovich2002,Zaslavsky2002461}, generalized Fokker-Planck equation \cite{Metzler20001,Zaslavsky2002461,Metzler1999431}, generalization of statistical mechanics (extensive and non-extensive) based on the Tsallis \cite{Essex2000299,Tsallis2001,Gell-Mann2004} and Renyj~\cite{Essex2000299,Vasconcellos20064821} entropy, and others were developed to describe the anomalous diffusion in different physical and chemical systems.
 Conducted researches show that
 mathematical basis of anomalous diffusion is equation with fractional derivatives \cite{Uchaikin2008500,Zaslavsky2002461}.
 In particular,
 during the study of three-dimensional models of anomalous diffusion~\cite{Uchaikin2008500,Uchaikin20131074,Uchaikin2003810},
 the basic equation of anomalous diffusion is derived
 from the general principles of the stochastic theory of random processes
 based on the Chapman-Kolmogorov integral equations for transition probability.
 Solutions of these equations form a new class of distributions, called fractional stable distributions.
 These distributions are solutions of partial differential equations of fractional order.
 These equations are generalization of usual diffusion equation to the case of anomalous diffusion.
 A partial case of the fractional stable distribution is the Gaussian distribution,
 which corresponds to the normal diffusion.
 It is important to note that the equations for anomalous diffusion with fractional derivatives
 contain diffusion coefficient, which is  a constant in time and space.
 On the other hand, the diffusion coefficients are related to time correlation functions
 (the Green-Kubo relations)
 containing diffusion transfer mechanisms from the perspective of nonequilibrium statistical mechanics.

 Currently,
 together with the phenomenological approach for constructing of the Fokker–Planck equation,
 the diffusion equation and its generalization --- the Cattaneo equation with fractional derivatives,
 there are two methods of constructing such equations,
 namely,
 (1)~probabilistic method,
 based on the Chapman-Kolmogorov equation in stochastic theory of random processes~\cite{Uchaikin2008500,Zaslavsky2002461,Stanislavsky2004418},
 and (2)~statistical method,
 based on the Liouville equation with fractional
 derivatives~\cite{Tarasov2004123,Tarasov200517,Tarasov2005011102,Tarasov2006033108,Tarasov2006341,Tarasov2013663,Tarasov20082984,Tarasov2005286,Tarasov2007237,Tarasov2007163,Tarasov2009179,Tarasov20121719,Tarasov2013102110,Tarasov2010}.
 In particular,
 by using this method,
 the BBGKY hierarchy equations with fractional derivatives~\cite{Tarasov200517,Tarasov2005011102,Tarasov2007237},
 transport equation,
 diffusion equation,
 and the Heisenberg equation with fractional derivatives~\cite{Tarasov2006341,Tarasov2013663,Tarasov20082984} are obtained.
 This approach is formulated for non-Hamiltonian systems.
 If the Helmholtz conditions for the coordinate and momentum derivatives are fulfilled,
 the Hamiltonian systems with the time-reversible Liouville equation with fractional derivatives are obtained
 from non-Hamiltonian systems.
 In Ref.~\cite{Kobelev20000002002},
 time-irreversible equations of motion of Hamilton and Liouville
 for dynamic of classical particles in space with multifractal time are offered.
 By using the definition of fractional derivative and the Riemann-Liouville integral,
 the time-irreversible Liouville equation with fractional derivatives
 (where the time is given on multifractal sets with fractional dimensions) is obtained.
 In Refs.~\cite{Kobelev2000194,Kobelev2002580},
 kinetic equation for systems with fractal structure
 (in particular, for description of diffusion processes in space of coordinates and momenta) is obtained
 within the Klimontovich approach.
 A similar approach for constructing of time fractional generalization
 for the Liouville equation and the Zwanzig equation (within projection formalism) is proposed in Ref.~\cite{Lukashchuk2013740}.

 In the present work,
 by using the Zubarev nonequilibrium statistical operator method~\cite{Zubarev19811509,Zubarev20021,Zubarev20022,Markiv2011785} and the maximum entropy principle for the Renyi entropies,
 we consider a way of obtaining generalized (non-Markovian) diffusion equation with fractional derivatives.
 Using the Liouville equation with fractional derivatives proposed by Tarasov In Refs.~\cite{Tarasov2004123,Tarasov200517,Tarasov2005011102,Tarasov2006033108}
 is an important and fundamental step for obtaining this equation.
 In the second section,
 by using the Zubarev nonequilibrium statistical operator method and
 the maximum entropy principle for the Renyi entropies,
 we found a solution of the Liouville equation with fractional derivatives at a selected set of observed variables.
 In the third section, we chose nonequilibrium average values of particle density as a parameter of reduced description,
 and then we received a generalized (non-Markovian) diffusion equation with fractional derivatives.
 
 \newpage

\section{Liouville equation with fractional derivatives for classical system of particles}

 We use the Liouville equation with fractional derivatives
 obtained by Tarasov in Refs.~\cite{Tarasov2004123,Tarasov200517,Tarasov2005011102,Tarasov2006033108}
 for non-equilibrium particle function $\rho(x^{N};t)$ of classical system:
 \begin{equation}\label{eq:2.1}
  \frac{\partial }{\partial t}\rho(x^{N};t)+\sum^{N}_{j=1}D^{\alpha}_{\vec{r}_{j}}
  \left(\rho(x^{N};t)\vec{v}_{j}\right)+\sum^{N}_{j=1}D^{\alpha}_{\vec{p}_{j}}
  \left(\rho(x^{N};t)\vec{F}_{j}\right)=0,
 \end{equation}
 where $x^{N}=x_{1},\ldots,x_{N}$,
 $x_{j}=\{\vec{r}_{j}, \vec{p}_{j}\}$ are the phase variables (coordinate and momentum) of $j$-th particle,
 $\vec{v}_{j}$ is the fields of velocity,
 $\vec{F}_{j}$ is the force field acting on $j$-th particle.
 \begin{equation}\label{eq:2.2}
  D^{\alpha}_{x} f(x)=\frac{1}{\Gamma (n-\alpha)} \int^{x}_{0} \frac{f^{n}(z)}{(x-z)^{\alpha +1-n}} dz
 \end{equation}
 is the Riemann-Liouville fractional derivative~\cite{Oldham2006,Samko1993},
 $n-1<\alpha<n$,
 $f^{n}(z)=\frac{d^{n}}{dz^{n}}f(z)$.
 In the general case we have
 \[
  D^{\alpha}_{\vec{r}_{j}}\left(\rho(x^{N};t)\vec{F}_{j}\right)
  \neq
  \rho(x^{N};t)D^{\alpha}_{\vec{r}_{j}}\vec{F}_{j}
  + \vec{F}_{j}D^{\alpha}_{\vec{r}_{j}}\rho(x^{N};t).
 \]
 If $\vec{F}_{j}$ does not depend on $\vec{p}_{j}$,
 and $\vec{v}_{j}$ does not depend on $\vec{r}_{j}$,
 we get
 \[
  \frac{\partial }{\partial t}\rho(x^{N};t)
  +
  \sum^{N}_{j=1}\vec{v}_{j}D^{\alpha}_{\vec{r}_{j}}\rho(x^{N};t)+\sum^{N}_{j=1}\vec{F}_{j}D^{\alpha}_{\vec{p}_{j}}
  \rho(x^{N};t)=0,
 \]
 \[
  \vec{v}_{j}=D^{\alpha}_{\vec{p}_{j}}H(\vec{r},\vec{p}),  \vec{F}_{j}=-D^{\alpha}_{\vec{r}_{j}}H(\vec{r},\vec{p}) ,
 \]
 where $H(\vec{r},\vec{p})$ is a Hamiltonian of system with fractional derivatives.
 We get the Liouville equation in the form
 \begin{equation}\label{eq:2.3}
  \frac{\partial }{\partial t}\rho(x^{N};t)
  +
  \sum^{N}_{j=1}D^{\alpha}_{\vec{p}_{j}}H(\vec{r},\vec{p})D^{\alpha}_{\vec{r}_{j}}\rho(x^{N};t)
  -
  \sum^{N}_{j=1}D^{\alpha}_{\vec{r}_{j}}H(\vec{r},\vec{p})D^{\alpha}_{\vec{p}_{j}}\rho(x^{N};t)
  =0,
 \end{equation}
 or
 \begin{equation}\label{eq:2.4}
  \frac{\partial }{\partial t}\rho(x^{N};t)+iL_{\alpha}\rho(x^{N};t)=0,
 \end{equation}
 where $iL_{\alpha}$ is the Liouville operator with the fractional derivatives,
 \begin{equation}\label{eq:2.5}
  iL_{\alpha}\rho(x^{N};t)
  =
  \left(\sum^{N}_{j=1}D^{\alpha}_{\vec{p}_{j}}H(\vec{r},\vec{p})D^{\alpha}_{\vec{r}_{j}}
  -
  \sum^{N}_{j=1}D^{\alpha}_{\vec{r}_{j}}H(\vec{r},\vec{p})D^{\alpha}_{\vec{p}_{j}}\right)\rho(x^{N};t).
 \end{equation}

 A solution the Liouville equation (\ref{eq:2.5}) will be found
 with the Zubarev nonequilibrium statistical operator method~\cite{Zubarev19811509,Zubarev20021}.
 After choosing parameters of the reduced description,
 taking into account projections
 we present the non-equilibrium particle function $\rho\left(x^{N};t\right)$
 (as a solution of the Liouville equation) in general form
 \begin{equation}\label{eq:2.6}
   \rho(x^{N};t)
   =
   \rho_{rel}\left(x^{N};t\right)
   -
   \int^{t}_{-\infty}e^{\varepsilon(t'-t)}T(t,t')(1-P_{rel}(t'))iL_{\alpha}\rho_{rel}(x^{N};t')dt',
 \end{equation}
 where
 $T(t,t')=\exp_{+}\left(-\int^{t}_{t'}(1-P_{rel}(t'))iL_{\alpha}dt'\right)$
 is the evolution operator containing the projection,
 $\exp_+$ is ordered exponential,
 $\varepsilon\to+0$ after taking the thermodynamic limit,
 $P_{rel}(t')$ is the generalized Kawasaki-Gunton projection operator
 depended on a structure of the relevant statistical operator (distribution function), $\rho_{rel}(x^{N};t')$.
 By using the Zubarev nonequilibrium statistical operator method~\cite{Zubarev19811509,Zubarev20021,Zubarev20022} and approach~\cite{Markiv2011785},
 $\rho_{rel}\left(x^{N};t'\right)$ will be found from the extremum of the Renyi entropy
 at fixed values of observed values $\langle \hat{P}_{n}(x)\rangle^{t}_{\alpha}$,
 taking into account the normalization condition $\langle 1 \rangle^{t}_{\alpha, rel}=1$,
 where the nonequilibrium average values are found respectively~\cite{Tarasov2005082106,Tarasov2006052107,Tarasov2005011102,Tarasov2006033108}
 \begin{equation}\label{eq:2.7}
  \langle \hat{P}_{n} (x)\rangle^{t}_{\alpha}=\hat{I}^{\alpha}(1,\ldots,N)\hat{T}(1,\ldots,N)\hat{P}_{n}\rho(x^{N};t).
 \end{equation}
 $\hat{I}^{\alpha}(1,\ldots,N)$ has the following form for a system of $N$ particles
 \[
  \hat{I}^{\alpha}(1,\ldots,N)
  =
  \hat{I}^{\alpha}(1),\ldots,\hat{I}^{\alpha}(N),
  \quad
  \hat{I}^{\alpha}(j)=\hat{I}^{\alpha}(\vec{r}_{j})\hat{I}^{\alpha}(\vec{p}_{j})
 \]
 and defines operation of integration
 \begin{equation}\label{eq:2.8}
  \hat{I}^{\alpha}(x)f(x)=\int^{\infty}_{-\infty}f(x)d\mu_{\alpha}(x),
  \quad
  d\mu_{\alpha}(x)=\frac{|x|^{\alpha}}{\Gamma (\alpha)}dx.
 \end{equation}
 The operator $\hat{T}(1,\ldots,N)=\hat{T}(1),\ldots,\hat{T}(N)$ defines the operation
 \[
  \hat{T}(x_{j})f(x_{j})=\frac{1}{2} \left(f(\ldots,x'_{j}-x_{j},\ldots)+f(\ldots,x'_{j}+x_{j},\ldots)\right).
 \]
 Accordingly,
 the average value,
 calculated with the relevant distribution function,
 is defined as
 \[
 \langle (\ldots) \rangle^{t}_{\alpha, rel}=\hat{I}^{\alpha}(1,\ldots,N)\hat{T}(1,\ldots,N)(\ldots)\rho_{rel}(x^{N};t).
 \]
 According to~\cite{Markiv2011785},
 the relevant distribution function has the form
 \begin{equation}\label{eq:2.9}
  \rho_{rel}(t)
  =
  \frac{1}{Z_{R}(t)}
  \left(1
  -
  \frac{q-1}{q}
  \beta
  \left(
  H-\sum_{n}\int d\mu_{\alpha}(x)F_{n}(x;t)\delta \hat{P}_{n}(x;t)
  \right)
  \right)^{\frac{1}{q-1}},
 \end{equation}
 where $Z_{R}(t)$ is the partition function of the Renyi distribution,
 which is determined from the normalization condition and has the form
 \begin{align}\label{eq:2.10}
   Z_{R}(t)&=\hat{I}^{\alpha}(1,\ldots,N)\hat{T}(1,\ldots,N) \nonumber\\
           &\quad \times
            \left(
            1-\frac{q-1}{q}\beta\left(H-
 \sum_{n}\int d\mu_{\alpha}(x)F_{n}(x;t)\delta \hat{P}_{n}  (x;t)\right)
            \right)^{\frac{1}{q-1}}.
 \end{align}
 The parameters $F_{n}(x;t)$ are determined from the self-consistency conditions
 \begin{equation}\label{eq:2.11}
   \langle \hat{P}_{n}(x) \rangle^{t}_{\alpha}= \langle \hat{P}_{n}(x) \rangle^{t}_{\alpha, rel}.
 \end{equation}

 In the next section we will consider a specific example of diffusion processes in dense gases and liquids.

\section{Generalized diffusion equation with fractional derivatives}

 The nonequilibrium particle number density
 $n(\vec{ r}_{\alpha};t)=\langle\hat n(\vec{r})\rangle^{t}_{\alpha}$
 (where $n(\vec{r})=\sum_{j=1}^{N}\delta(\vec r-\vec r_{j})$ is the microscopic particle number density)
 is the main parameter of the reduced description
 for describing diffusion processes in classical gases and liquids.
 For such choice of parameter of the reduced description,
 the relevant distribution function has the form
 \begin{equation}\label{eq:2.12}
  \rho_{rel}(t)
  =
  \frac{1}{Z_{R}(t)}
  \left(
  1-\frac{q-1}{q}\beta
  \left(H-\int d\mu_{\alpha}(\vec r)\nu (\vec r;t)\delta\hat n (\vec r_{\alpha};t)\right)
  \right)^{\frac{1}{q-1}},
 \end{equation}
 where
 \begin{equation}\label{eq:2.13}
  Z_{R}(t)=\hat{I}^{\alpha}(1,\ldots,N)\hat{T}(1,\ldots,N)
  \left(1
  -\frac{q-1}{q}\beta
  \left(H-
  \int d\mu_{\alpha}(\vec r)\nu (\vec r;t)\delta\hat n (\vec r_{\alpha};t)\right)
  \right)^{\frac{1}{q-1}}
 \end{equation}
 is the partition function of the relevant distribution function,
 $\beta=\frac{1}{k_{\text{B}}T}$,
 $k_{\text{B}}$ is the Boltzmann constant,
 $T$ is the equilibrium value of temperature,
 $\delta\hat n(\vec r_{\alpha};t)=\hat n(\vec r)-\left\langle\hat n(\vec r)\right\rangle^{t}_{\alpha}$
 is the fluctuations of the density,
 and parameter $\nu(\vec r;t)$ is determined by the self-consistency condition
 \begin{equation}\label{eq:2.14}
  \left\langle \hat n(\vec{ r})\right\rangle^{t}_{\alpha}=\left\langle \hat n(\vec{ r})\right\rangle^{t}_{\alpha, rel}.
 \end{equation}
 It should be noted
 that the relevant distribution function (\ref{eq:2.12}) within the Renyi statistics for $q=1$
 coincides with one within the Gibbs statistics~\cite{Zubarev19811509}.
 The distribution (\ref{eq:2.12}) can be represented as
 \begin{equation}\label{eq:2.15}
   \rho_{rel}(t)
   =
   \frac{1}{Z_{R}(t)}
   \left(
    1-\frac{q-1}{q}\beta
    \left(
      H-\int d\mu_{\alpha}(\vec r)\nu^{*} (\vec r;t)\hat n (\vec r)
    \right)
   \right)^{\frac{1}{q-1}},
 \end{equation}
 where
 \begin{equation}\label{eq:2.16}
  Z_{R}(t)=\hat{I}^{\alpha}(1,\ldots,N)\hat{T}(1,\ldots,N)
  \left(
  1-\frac{q-1}{q}\beta
  \left(H-\int d\mu_{\alpha}(\vec r)\nu^{*} (\vec r;t)\hat n (\vec r)\right)
  \right)^{\frac{1}{q-1}},
 \end{equation}
 \[
   \nu^{*}(\vec r;t)
   =
   \frac{\displaystyle\nu(\vec r;t)}
   {\displaystyle 1+\frac{q-1}{q}\int d\mu_{\alpha}(\vec r) \nu(\vec r;t)\left\langle\hat n(\vec r)\right\rangle^{t}_{\alpha}}.
 \]
 Substituting Eq.~(\ref{eq:2.15}) in Eq.~(\ref{eq:2.6}),
 we find the nonequilibrium statistical operator
 \begin{equation}\label{eq:2.17}
  \rho(t)=\rho_{rel}(t)
  +
  \int_{-\infty}^{t}e^{\varepsilon(t'-t)}
  T(t,t')
  \int d\mu_{\alpha}(\vec r')I_{n}(\vec r'_{\alpha};t')\rho_{rel}(t)\nu^{*}(\vec r';t)dt',
 \end{equation}
 where
 \begin{equation}\label{eq:2.18}
  I_{n}(\vec r_{\alpha};t)=\left(1-P(t)\right)\frac{1}{q}\psi^{-1}(t)iL_{\alpha}\hat n(\vec r)
 \end{equation}
 is the generalized flow,
 \[
  \psi(t)=1-\frac{q-1}{q}\beta\left(H-\displaystyle\int d\mu_{\alpha}(\vec r)\nu^{*}(\vec    r;t)\hat n(\vec r)\right).
 \]
 $P(t)$ is the projection operator that has the following structure:
 \[
  P(t)\ldots=\int d\mu_{\alpha}(\vec r) \int d\mu_{\alpha}(\vec r')
  \left\langle\ldots\hat n (\vec r)\right\rangle_{\alpha, rel}^{t}
  [\langle\hat n(\vec r)\delta\{[q\psi(t)]^{-1}\hat n(\vec r')\}\rangle_{\alpha,rel}^{t}]^{-1}
  \delta\{[q\psi(t)]^{-1}\hat n (\vec r')\},
 \]
 where
 $\delta\{A\}=A-\langle A\rangle_{\alpha, rel}^{t}$.

 By using the nonequilibrium statistical operator (\ref{eq:2.17}),
 we get the generalized diffusion equation
 for the parameter of the reduced description
 \begin{equation}\label{eq:2.19}
  \frac{\partial}{\partial t}\left\langle\hat n(\vec r)\right\rangle^{t}_{\alpha}=
  \int d\mu_{\alpha}(\vec r')
  \int_{-\infty}^{t}e^{\varepsilon(t'-t)}\phi_{nn}(\vec r, \vec r';t,t')\beta\nu^{*}(\vec r';t')dt',
 \end{equation}
 where
 \begin{align*}
   \varphi_{nn}(\vec r,\vec r';t,t')&=\hat{I}^{\alpha}(1,..,N)\hat{T}(1,..,N)iL_{\alpha}\hat n(\vec r)T(t,t')I_{n}(\vec r'_{\alpha};t')\rho_{rel}(x^{N};t')\\
    & =\frac{\partial^{\alpha} }{\partial \vec r^{\alpha}}\cdot D_{q}(\vec r,\vec           r';t,t') \cdot\frac{\partial^{\alpha} }{\partial\vec r^{'\alpha}}
 \end{align*}
 is the generalized transport kernel (memory function),
 in which the averaging is performed with the distribution function~(\ref{eq:2.15}).

 As a result,
 we get the non-Markovian diffusion equation with fractional derivatives
 \begin{equation}\label{eq:2.20}
   \frac{\partial}{\partial t}\left\langle\hat n(\vec r)\right\rangle^{t}_{\alpha}=\frac{\partial^{\alpha} }{\partial\vec r^{\alpha}}\cdot\displaystyle\int d\mu_{\alpha}(\vec r')\displaystyle\int_{-\infty}^{t}e^{\varepsilon(t'-t)}D_{q}(\vec r,\vec r';t,t')\cdot\frac{\partial^{\alpha} }{\partial\vec r ^{'\alpha}}\beta\nu^{*}(\vec r';t')dt',
 \end{equation}
 where
 \begin{align}\label{eq:2.21}
  D_{q}(\vec r, \vec r';t,t')&=\langle\hat{\vec{v}}(\vec r)T(t,t')\hat{\vec{v}}(\vec r')\rangle^{t}_{\alpha, rel}\nonumber\\
  &=\hat{I}^{\alpha}(1,\ldots,N)\hat{T}(1,\ldots,N)\nonumber\\
  &\quad\times\hat{\vec{v}}(\vec r)T(t,t')\hat{\vec{v}}\vec (r')
    \frac {1}{Z_{R}(t)}
    \left(1-\frac{q-1}{q}\beta\left(H-\int d\mu_{\alpha}(\vec r)\nu^{*}(\vec r;t)\hat n(\vec r)\right)\right)^{\frac{1}{q-1}}
 \end{align}
 is the generalized diffusion coefficient within the Renyi statistics,
 in which the averaging is performed with the distribution function (\ref{eq:2.15}),
 where $\hat{\vec{v}}(\vec r)=\sum_{j=1}^{N}\vec v_{j}\delta(\vec r-\vec r_{j})$
 is the microscopic particle number flux density.
 At $q=1$
 the generalized diffusion equation within the Renyi statistics goes over to
 one within the Gibbs statistics with fractional derivatives.
 If $q=1$ and $\alpha=1$,
 we get the generalized diffusion equation within the Gibbs statistics~\cite{Zubarev19811509}.
 In the Markov approximation for the general diffusion coefficient in time and space
 $D_{q}(\vec r, \vec r';t,t')\approx D_{q}\delta (t-t')\delta (\vec r -\vec r')$,
 after exclusion the parameter $\nu^{*}(\vec r';t')$ by using the self-consistency condition,
 from Eq.~(\ref{eq:2.20}) we obtain the diffusion equation with fractional derivatives
 \begin{equation} \label{eq:2.22}
  \frac{\partial}{\partial t}\left\langle\hat n(\vec r)\right\rangle^{t}_{\alpha}
  =D_{q} \frac{\partial^{2\alpha} }{\partial r^{2\alpha}}
  \left\langle\hat n(\vec r)\right\rangle^{t}_{\alpha}.
 \end{equation}

 \section{Conclusion}

 By using the Zubarev nonequilibrium statistical operator~\cite{Zubarev19811509,Zubarev20021,Zubarev20022,Markiv2011785},
 the Liouville equation with fractional derivatives~\cite{Tarasov2005082106,Tarasov2006052107,Tarasov2005011102,Tarasov2006033108},
 and the principle of maximum entropy,
 we obtain the generalized (non-Markovian) diffusion equation with fractional derivatives.
 By using this approach,
 the generalized transfer equation with fractional derivatives can be obtained
 at some set of parameters of the reduced description $\langle\hat{P}_{n}(x)\rangle ^ {t}_{\ alpha}$
 of nonequilibrium state of the system
 In particular,
 if these parameters are nonequilibrium average values of particle density, momentum and energy,
 we obtain generalized hydrodynamic equations with fractional derivatives,
 which are generalization of Tarasov's results \cite{Tarasov2005286}.

\bibliography{MyBasa2}%
\bibliographystyle{apsrev} 

\end{document}